\documentclass[fleqn,usenatbib]{mnras}

\usepackage{newtxtext,newtxmath}

\usepackage[T1]{fontenc}
\usepackage{ae,aecompl}

\usepackage{graphicx}	
\usepackage{amsmath}	
\usepackage{amssymb}	

\newcommand\lsim{\mathrel{\rlap{\lower4pt\hbox{\hskip1pt$\sim$}}
        \raise1pt\hbox{$<$}}}
\newcommand\gsim{\mathrel{\rlap{\lower4pt\hbox{\hskip1pt$\sim$}}
        \raise1pt\hbox{$>$}}}

\title[Late inspiral of a SMBHB with a circumbinary disc]{The late inspiral of supermassive black hole binaries with circumbinary gas discs in the LISA band}

\author[Y. Tang, Z. Haiman and A. MacFadyen]{
Yike Tang$^{1}$,
Zolt\'{a}n Haiman$^{2}$,
Andrew MacFadyen$^{1}$\thanks{E-mail: yt611@nyu.edu~(YT); zoltan@astro.columbia.edu (ZH); macfadyen@nyu.edu~(AM)}
\\
$^{1}$Center for Cosmology and Particle Physics, Physics Department, New York University, New York, NY, USA,10003\\
$^{2}$Department of Astronomy, Columbia University, New York, NY, USA,10027\\
}

\date{Accepted XXX. Received YYY; in original form ZZZ}

\pubyear{2018}

\begin{document}
\label{firstpage}
\pagerange{\pageref{firstpage}--\pageref{lastpage}}
\maketitle

\begin{abstract}
We present the results of 2D, moving-mesh, viscous hydrodynamical
simulations of an accretion disc around a merging supermassive black
hole binary (SMBHB). The simulation is pseudo-Newtonian, with the BHs
modeled as point masses with a Paczynski-Wiita potential, and includes
viscous heating, shock heating, and radiative cooling.  We follow the
gravitational inspiral of an equal-mass binary with a component mass
$M_{\rm bh}=10^{6}M_{\odot}$ from an initial separation of $60r_{\rm
  g}$ (where $r_{\rm g}\equiv GM_{\rm bh}/c^{2}$ is the gravitational
radius) to the merger.  We find that a central, low-density cavity
forms around the binary, as in previous work, but that the BHs capture
gas from the circumbinary disc and accrete efficiently via their own
minidiscs, well after their inspiral outpaces the viscous evolution of
the disc.  The system remains luminous, displaying strong periodicity
at twice the binary orbital frequency throughout the entire inspiral
process, all the way to the merger.  In the soft X-ray band, the
thermal emission is dominated by the inner edge of the circumbinary
disc with especially clear periodicity in the early inspiral.  By
comparison, harder X-ray emission is dominated by the minidiscs, and
the light curve is initially more noisy but develops a clear
periodicity in the late inspiral stage.  This variability pattern
should help identify the EM counterparts of SMBHBs detected by the
space-based gravitational-wave detector LISA.
\end{abstract}




\section{Introduction}

Supermassive black holes (SMBH) are currently believed to reside in
most, if not all galactic nuclei, and SMBH binaries (SMBHBs) are
expected to be produced frequently in mergers of galaxies
(e.g. \citealt{KormendyHo2013}). The resulting compact SMBHBs are
likely embedded in a gaseous environment. Electromagnetic (EM)
signatures of such compact SMBHBs can arise from their interaction
with gas.

The space-based gravitational-wave (GW) detector LISA will be
sensitive to SMBHBs with masses in the range $10^{4}-10^{7}~{\rm
  M_{\odot}}$ \citep{LISA}. Identifying a GW source in the EM bands
has considerable importance for astrophysics and cosmology
(e.g. \citealt{2009astro2010S.235P}). It has been proposed that
periodic emission in the EM bands may track the orbital motion of
SMBHBs throughout the late inspiral, and that this would then allow a
unique identification of a LISA source~\citep{Kocsis2008}.  Comparing
the EM and GW chirp signals would then also help probe the difference
in the propagation speed of photons and gravitons
\citep{Kocsis2008,Haiman2017}, complementing measurements of the
graviton speed from phasing of the GWs alone (see, e.g., the review by
\citealt{Will2006} and references therein).  A similar constraint has
recently been derived from the time-delay between the GWs and the
accompanying gamma-ray burst arriving from the neutron-star merger
GW170817~\citep{ligo-bns}.  A SMBHB offers an improved and more robust
measurement of this time-delay; this is especially the case if Doppler
modulations of the EM chirp are detected from the orbital motion of
the BHs themselves, fixing the relative phase of the EM and GW signals
\citep{Haiman2017}.

In order to predict the EM signatures of SMBHB mergers in the LISA
band, the first step is to resolve the gas dynamics during this late
inspiral stage. The morphology of the gas, even on scales as large as
that of the circumbinary disc, is sensitive to the behavior of gas
streams in the vicinity of the individual BHs \citep{Yike17,Bowen+2017}. One
necessary requirement for reliably simulating the LISA stage is to
resolve the accretion flows near the individual BHs, on scales down to
that of the innermost stable circular orbit (ISCO; located at
$6GM_{\rm bh}/c^2\equiv 6r_{\rm g} $ for a non-spinning BH).
Furthermore, the gas dynamics can not be in steady state, since the
binary orbit is evolving rapidly.  Depending on the component masses,
for a circular, comparable-mass binary, the LISA band corresponds to
the inspiral from an initial separation of $(50-200)GM/c^{2}$ all the
way to the merger, and covers the last $\approx100$ to a few thousand
orbits of the binary.  The orbital evolution of the binary is likely
overwhelmingly dominated by its gravitational radiation at this late
stage \citep{HKM09,Dotti:2012:rev,Yike17}.

Because of the dual requirements of high spatial resolution and large
number of orbits, simulating the LISA stage in its entirety has
remained beyond the capability of fully three-dimensional (GR)MHD
simulations. They have generally focused instead on the last
$\approx$10 orbits preceding the merger \citep{Bode+2010,Noble+2012,
  Farris2012,Shi+2012, Giacomazzo2012,bowen17}.  {\em However,
  following the binary throughout the entire LISA band is feasible in
  2D hydrodynamics}.  Previous high-resolution Newtonian simulations
have indeed followed binaries for thousands of orbits. However, these
works have focused on large separations, well before the LISA stage
\citep{2015MNRAS.446L..36F,2012A&A...545A.127R,2009MNRAS.393.1423C,Krolik+2010}. As
such, they did not directly resolve accretion inside the ISCO
(typically either ignoring accretion, or implementing accretion via a
sink prescription).

In our earlier work \citep{Farris15}, we performed a 2D simulation of
an inspiraling and merging BH binary, embedded in viscous circumbinary
disc. However, this earlier study was purely Newtonian, and employed
an artificial sink that depleted the gas according to the fluid
viscous time scale. Moreover, the focus was on the overall accretion
rate, and we did not present an analysis of the gas dynamics and did
not compute the emerging EM emission and light-curves.

In this paper we focus on modeling the entire inspiral stage
observable with LISA, extending our earlier work.  We consider an
equal-mass, non-spinning BH binary with a component mass of $M_{\rm
  bh}=10^{6}M_{\odot}$ and initial separation $60 r_{\rm g}$.  This
corresponds to the last $\approx 100$ orbits of the merger, or the
last $\approx(1+z)/2$ week observed for a binary at redshift~$z$.
The binary is assumed to be on a circular orbit, and we allow the
orbit to shrink due to GW emission all the way to merger. The binary
is embedded in a viscous, optical thick $\alpha$-disc with radiative
cooling. We employ a pseudo-Newtonian potential for the black
holes. Our spatial resolution is sufficient to resolve the individual
BH's ISCO, allowing us to remove gas inside the event horizon
physically. We analyse the dynamics and the time-dependent spectral
signatures of the emerging thermal emission.

This paper is organized as follows. In \S~\ref{sec:setup}, we
summarize our numerical methods. In \S~\ref{sec:results}, we present
our main findings, including disc spectra and light curves. We also
examine the periodicities of the light curves separately in the soft
and hard X-ray bands. In \S~\ref{sec:discuss}, we discuss the
implications of our results, along with some caveats. Finally, in
\S~\ref{sec:conclude} we summarize our conclusions and indicate topics
for future work.

\vspace{-1\baselineskip}
\section{Numerical Setup}
\label{sec:setup}

The model and numerical setup of this work closely resembles that in
our previous work \citep{Farris15}. We briefly outline the key
aspects here and highlight the modifications we made for the present
study. We refer the reader to the above paper for further details.

All simulations are performed using the publicly available moving-mesh
grid code \texttt{DISCO} in 2D \citep{Duffell+2012,DISCOrelease}.  We
simulate an equal-mass SMBHB with $M_{\rm bin}=2M_{\rm bh}=2\times
10^{6}M_{\odot}$ and initial separation $a_0=60GM_{\rm bh}/c^{2}$.  At
this separation, a typical $10^6 {\rm M_\odot}$ binary at $z<1$ can be
localized by LISA on the sky to a few square degrees, giving the
chance to a wide-field telescope to locate and identify the source
\citep{Kocsis2008,Lang2008}. We assume the BHs are non-spinning so
that $a_0=10r_{\rm isco}=30r_{\rm S}$, where $r_{\rm S}$ is the
Schwarzschild radius of an individual BH.  The BHs are placed on the
grid, and the vertically-integrated fluid equations are solved
assuming an $\alpha$-viscosity with $\alpha=0.1$.

Inside the cavity that forms around the binary (with a size roughly
twice the initial binary separation), the grid spacing is roughly
linear in radius and $r_{\rm isco}$ is resolved with $\approx$8 radial
cells; there are $\approx$ 100 cells in total inside the ISCO.
Outside the cavity, the radial grid spacing increases approximately
logarithmically up to 30$a_0$, allowing us to concentrate
computational resources on the flow in the inner regions of the disc.
The total number of radial cells is 512.  At all radii, we chose our
azimuthal grid spacing such that the aspect ratio of each cell is kept
approximately unity, unless this requires a number of azimuthal grid
cells $N_{\phi}>512$, in which case we cap the number of $\phi$-zones
and set $N_{\phi}=512$.

To mimic the effect of black hole accretion, gas inside the event
horizon (roughly the innermost 2 or 3 radial cells around each BH) is
depleted effectively instantaneously at each time-step using a sink
prescription similar to \cite{Yike17}. We use a sink radius $r_{\rm
  sink} = r_{\rm S}$ and a sink time scale $\approx 10^{-4}$ orbital
period. In this paper we assume the mass and orbit of the binary black
hole are not affected by the accretion on the time scale of our
simulations.

Our simulation is pseudo-Newtonian, i.e. we model the black holes as point
masses with the modified potential~\citep{1980A&A....88...23P}
\begin{equation}
\psi(r)=-GM/(r-r_{\rm S}).
\end{equation}
Since the Paczynski-Wiita potential diverges at $r=r_{\rm S}$, we
employ an upper limit for the gravitational force:
$|f(r)|=|f(1.5r_{\rm S})|$ for $ r<1.5r_{\rm S}$. We choose $1.5r_{\rm
  S}$ because our cell size is $\approx 0.5r_{\rm S}$.

We use a $\Gamma$-law equation of state for the gas in the form
$P=(\Gamma-1)e$, where P and $e$ are the vertically integrated
pressure and internal energy, respectively, and $\Gamma=5/3$ is the
adiabatic index. Radiative cooling is incorporated naturally through
the energy equation, as described in \cite{2015MNRAS.446L..36F}. We
assume the disc is optically thick and geometrically thin (the optical
depth due to electron scattering at $r\sim a$ near the time of
decoupling is $\sim 10^{5}$; \citealt{HKM09}).  These assumptions are
consistent with our simulation results.  The cooling rate is $q_{\rm
  cool}=4\sigma/3\tau T^{4}$, where $T$ is the mid-plane temperature.
Following \cite{2015MNRAS.446L..36F} and \cite{Farris15} we
parameterize our simulations by specifying the Mach number at the
fixed initial radius $r=a_0$,
$\mathcal{M}_{a}=v_{a}/\sqrt{P(a)/\Sigma(a)}$, where
$v_a\equiv(GM_{\rm bin}/a)^{1/2}$ and $\Sigma$ is the surface density,
We then obtain the local cooling rate from the scaling
\begin{equation}
  q_{\rm cool}=\frac{9}{8}\alpha\Omega(a)P(a)\mathcal{M}^{8}_{a}\left({\frac{P}{\Sigma v^{2}_a}}\right)^{4}\left(\frac{\Sigma}{\Sigma(a)}\right),
  \label{eq:qcool}
\end{equation}
where $\Omega$ is the Keplerian angular frequency\footnote{We note
  there is a typographical error in eq. 2 of
  \cite{2015MNRAS.446L..36F}; the correct exponent of the Mach number
  is $\mathcal{M}_a^{8}$, as in eq.~\ref{eq:qcool} above.}.  Changing
$\mathcal{M}_{a}$ is equivalent to changing the thickness $h/r$ of the
disc, and in this work we fix $\mathcal{M}_{a}=(h/r)^{-1}=10$. A
standard steady-state quasar disc \citep{SS1973} with a near-Eddington
accretion rate $\sim \dot{M}_{\rm Edd}$ has a scale-height of $h/r
\sim 10^{-3}$, and is much thinner than the disc in our simulations
($h/r \sim 0.1$). Simulating an accretion disc with $h/r \sim 10^{-3}$
is prohibitively challenging, and our choice is typical of values
employed in previous numerical works
(e.g. \citealt{2013MNRAS.436.2997D}, \citealt{2014ApJ...783..134F},
\citealt{Farris15}; \citet{2017MNRAS.466.1170M} and
\citet{Ragusa+2016} have included slightly smaller $h/r$ values
ranging from 0.02 to 0.13) and allows for a direct comparison with
those works. The optical thickness to electron scattering ($\tau_{\rm
  es}>10$) and free-free absorption ($\tau_{\rm ff}>10^{4}$) are both
very high in our simulation domain (see, e.g., Figure 10 in
\citealt{Corrales2010}). Therefore throughout this paper we compute
the spectrum assuming multi-color thermal body radiation
\begin{equation}
L_{\nu}=\int\frac{2h\nu^{3}}{c^{2}(\exp\left[\frac{h\nu}{kT_{\rm eff}(r,\phi)}\right]-1)}dA,
\end{equation}
where $T_{\rm eff}(r,\phi)$ is the effective temperature obtained from
the local cooling rate $q_{\rm cool}(r,\phi)$ in the cell at
$(r,\phi$) via $T_{\rm eff}=(q_{\rm cool}/\sigma)^{0.25}$ with
$\sigma$ the Stefan-Boltzmann constant.

While computing the observed spectrum and light curve, we additionally
consider the relativistic Doppler effect and the gravitational
redshift.  For simplicity, and to assess the maximal impact of the
Doppler effect, we further assume the circumbinary disc is observed
edge-on. Photons emitted by gas suffer a Doppler shift in frequency
$D=[\Gamma(1-\beta_{\parallel})]^{-1}$, where
$\Gamma=(1-\beta^{2})^{-1/2}$ is the Lorentz factor, and
$\beta_{\parallel}$ the component of the velocity along the line of
sight.  The apparent flux $F_\nu$ at a fixed observed frequency $\nu$
is modified from the flux of a stationary source $F_\nu^{\rm 0}$ to
$F_\nu =D^{3}F^{\rm 0}_{\nu/D}$.  For the gravitational redshift, we
employ $D=(1-r_{\rm S}/R_{\rm e})^{-0.5}$, where $r_{\rm S}$ is the
Schwarzschild radius and $R_{\rm e}$ is the distance between the
source of emission and the BH.  The above is an approximation to a
fully relativistic ray-tracing in the binary's dynamical metric
\citep{Schnittman+2017}. In particular, we ignore gravitational
self-lensing \citep{Haiman2017,DorazioDiStefano2017} and other
relativistic effects, such as light-travel time modulations
(i.e. Shapiro delay), which appear at order $v/c$.

Our disc has an initial surface density profile
$\Sigma(r)=\Sigma_{0}(r/a)^{-0.5}$ and contains an artificial cavity
with the density depleted inside the radius 2.5$a_0$, as in
\citet{Yike17}.  Viscosity in the circumbinary disc transports angular
momentum outward, causing the cavity to refill. We first hold the
binary on the initial circular orbit and evolve the system for several
hundred orbits to reach a quasi-steady state. We then allow the binary
separation to shrink due to quadrupole formula for gravitational wave
emission \citep{Peters1964},
\begin{equation}
a(t)=a_0(1-t/\tau)^{0.25}.
\end{equation}
Here $\tau=7.22$ days is the GW inspiral time from the initial
separation of $30~GM_{\rm bin}/c^2$.  Finally, we employ a set of
convenient code units, in which the initial binary period is
$T(0)=2\pi$ and $a_0=1$; as a result, $\tau=385.17$ in code units.

\section{Results}
\label{sec:results}

As stated above, before allowing the black holes to inspiral and
merge, we first hold the binary separation fixed, and evolve the
system for several hundred orbits to reach a quasi-steady state. This
quasi-steady state is similar to that in previous
works~\citep{2013MNRAS.436.2997D,2014ApJ...783..134F,2015MNRAS.446L..36F,Yike17}
in that the cavity becomes lopsided, and two narrow streams
periodically connect the minidiscs around the individual BHs with the
circumbinary disc. Similar to \cite{Farris15}, we find that these
features persist until right before the merger, with only a gradual
decline in the size and surface density of the minidiscs and the
streams.

\begin{figure*}
\includegraphics[scale=0.4]{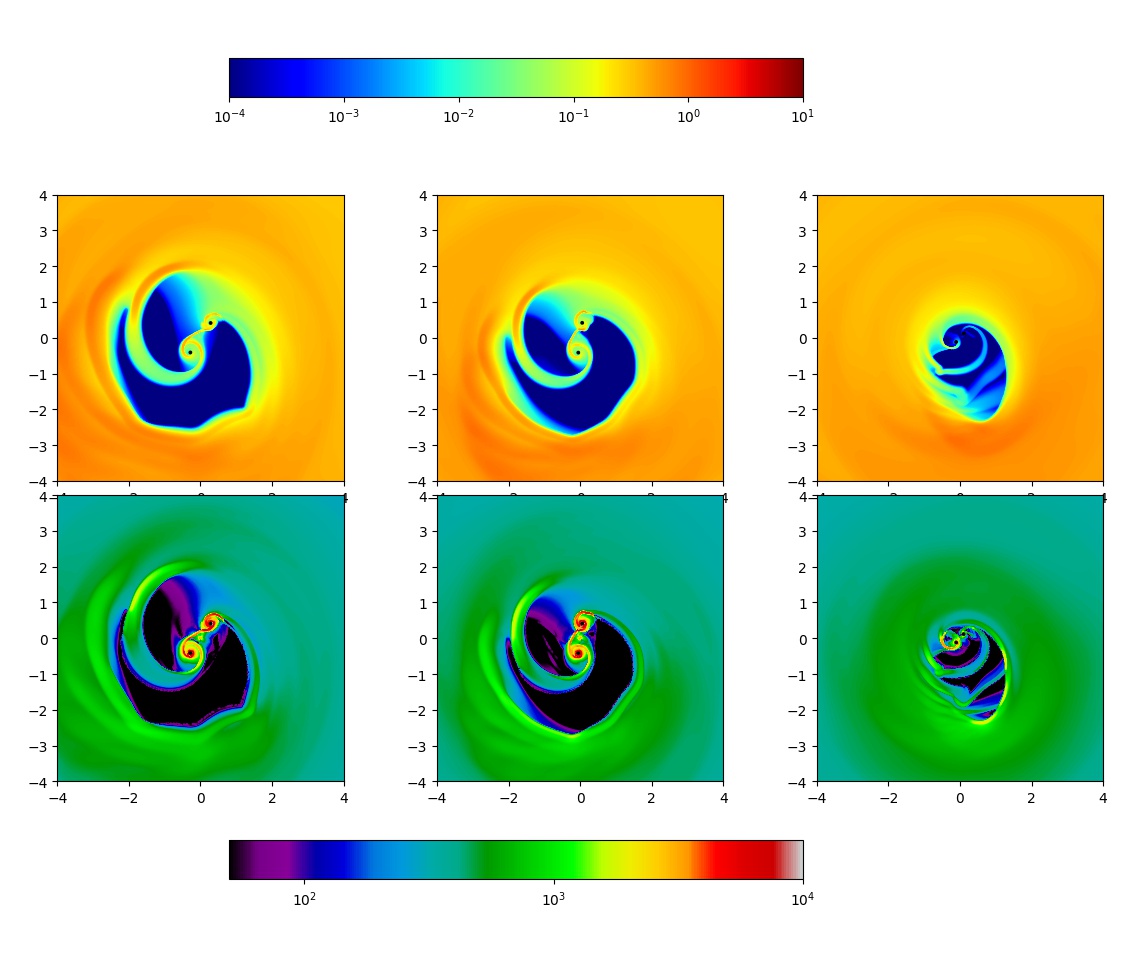}
\protect\caption{The top row shows snapshots of the surface density
  $\Sigma/\Sigma_0$, where $\Sigma_0$ is the initial surface density
  at $r=a$, and the bottom row shows snapshots of the effective
  temperature $T_{\rm eff}$ in [eV]. The horizontal and vertical axes
  are in units of the initial binary separation $a_0$.
  The snapshots are taken, from left to right, at the
  times $t=0.0\tau$, $0.5\tau$ and $0.99\tau$, where $\tau=7.22$days
  corresponds to the merger.  The cavity size decreases and the BHs
  accrete efficiently during the entire inspiral.}
\label{density_Teff}
\end{figure*}

\begin{figure}
  \includegraphics[scale=0.4]{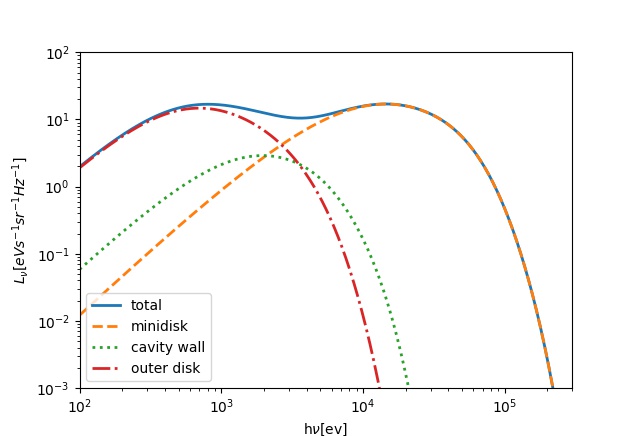}
  \protect\caption{Thermal spectrum computed from a simulation
    snapshot at the beginning of the inspiral ($t=0$). The full
    spectrum is shown by the blue curve. The distinct components
    arising from gas in the minidiscs ($r<a$), streams and cavity wall
    ($a<r<3a$) and from the outer regions of the circumbinary disc
    ($r>3a$) are represented by the orange dashed line, green dot line
    and red dash dot line, respectively.}
\label{spectra_decom}
\end{figure}

In Figure~\ref{density_Teff} we show the 2D distributions of surface
density and effective temperature at three different snapshots,
corresponding to $t=0$ (the beginning of the inspiral),
$0.5\tau$ (half-way to merger) and $0.99\tau$ (just before the
merger).
A spectrum computed at the beginning of the inspiral is also shown in
Figure~\ref{spectra_decom}. In this figure, we additionally show three
distinct components of the spectrum, arising from outer
disc~($r>3a_0$), the shocked streams inside the cavity and shocked hot
gas near the cavity wall ($a_0<r<3a_0$) and minidiscs ($r<a_0$).  At
frequencies below $\sim 1$keV, the emission from the viscously heated
outer disc dominates, while at higher frequencies, the emission mostly
comes from the shock-heated gas in the minidiscs, streams, and near
the cavity wall. At frequencies between $\sim$(1-20) keV, the total
spectrum shows a plateau with a small depression near $\sim$3 keV,
resembling the ``notch'' discussed in \citet{Roedig+2014}.

\begin{figure}
  \includegraphics[scale=0.41]{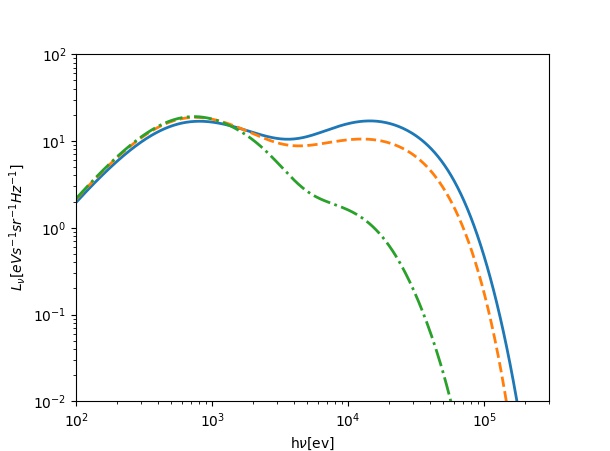}
  \protect\caption{Total composite thermal spectra computed from
    simulation snapshots at $t=0$ (blue curve), $t=0.8\tau$ (orange
    dashed) and $t=0.99\tau$ (green dot-dashed).}
\label{spectra}
\end{figure}

\begin{figure}
\includegraphics[scale=0.35]{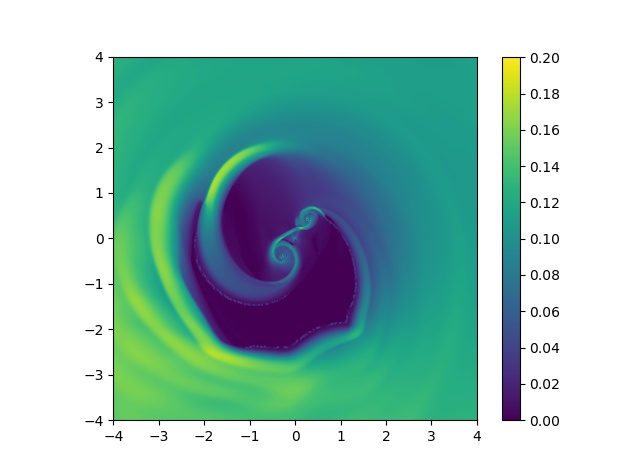}
\protect\caption{Local disc thickness $h/r$ throughout the disc at $t=0$.}
\label{hor}
\end{figure}

In Figure~\ref{spectra} we show the total spectrum at three different
snapshots, $t=0$ (blue; same as in Fig.~\ref{spectra_decom}),
$0.8\tau$ (orange) and $0.99\tau$ (green). The low-frequency part of
the spectrum, which is dominated by the outer regions of the
circumbinary disc, shows little change throughout the simulation. By
comparison, at higher frequencies ($\gsim 1$ keV), which arises from
the minidiscs or streams in the cavity, declines noticeably by
$0.8\tau$; the decline in the hard X-ray band becomes significant
during the end of inspiral ($0.99\tau$).  Note that in this paper we
chose $M_{\rm bin}=2\times 10^{6}{\rm M_{\odot}}$, so that the GW
signal is in the LISA band, and $0.99\tau$ corresponds to the last
$\approx2$ hours of the inspiral.  Note also that the photon frequency
scales with binary mass as $\nu\propto M_{\rm bin}^{-1/8}$.  This is
because we fixed $\mathcal{M}_{a}=10$, and the velocity is independent
of $M_{\rm bin}$.  Therefore $\Sigma_{0}$ scales as $M_{\rm bh}^{0.5}$
and $T_{\rm eff}$ scales as $M_{\rm bin}^{-1/8}$.  For a steady disc
around a solitary BH, it would further scale with Mach number as
$h\nu\propto \mathcal{M}_{\rm a}^{-5/4}$
($\Sigma\propto\mathcal{M}^{-3}$ and $T\propto\mathcal{M}^{-2}$).
This suggest that for a thinner disc, the frequencies could be lower
than shown for our fiducial $\mathcal{M}_{a}=10$ in
Figures~\ref{spectra_decom} and \ref{spectra}. However, for gas near
the cavity wall and minidiscs, the variation of the disc thickness
caused by its interaction with the binary BHs require further study
(e.g. \citealt{Ragusa+2016}).  In Figure~\ref{hor}, we show $h/r$ of
our simulated disc. In the shock heated region $h/r$ is increased to
$\sim 0.2$, while in the outer region $h/r$ stays at $\sim 0.1$.


\begin{figure}
  \hspace{-0.3in}\includegraphics[scale=0.41]{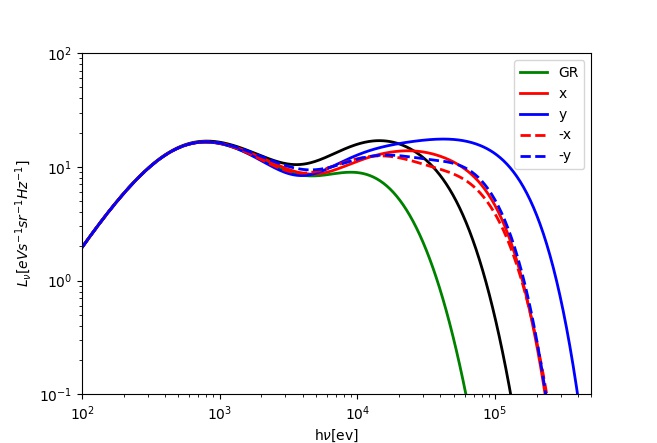}
  \protect\caption{Spectra illustrating the impact of the special
    relativistic Doppler shift and the gravitational red/blue
    shift. For reference, the black curve shows the spectrum without
    considering either of these effects. The green curve shows the
    spectrum including only the gravitational redshift, while the
    other four spectra, shown in blue and red, include both
    relativistic effects (for the four different viewing directions
    $\pm\hat{x}$ and $\pm\hat{y}$, as labeled in the inset).  All
    spectra are calculated from the snapshot at $t=0$, when the
    orbital velocity of the BHs is 0.091$c$.}
\label{spectra_doppler}
\end{figure}

In Figure~\ref{spectra_doppler}, we illustrate the effects of the
relativistic Doppler shift and gravitational redshift on the observed
spectra. As mentioned in \S~\ref{sec:setup}, we assume the disc is
observed edge-on, so that the Doppler effect is maximized. We include
four viewing angles in the plane of the disc, corresponding to the
$\pm\hat{x}$ and $\pm\hat{y}$ directions. As the figure shows, at the
low-frequency end ($h\nu\lsim 1$ keV), the spectrum is not modified by
either the Doppler or gravitational shifts.  A higher frequencies (1
keV $\lsim h\nu\lsim20$ keV) the Doppler shift causes an overall
dimming (i.e. red and blue curves), while at the highest frequencies
($h\nu\gsim20$ keV), the spectrum is brightened significantly.
Interestingly, we find that the dimming and the brightening retain
their sign in all four different viewing directions; we offer a simple
interpretation of this result in \S~\ref{sec:discuss} below.  For
reference, the figure also shows a case when only the GR redshift
(green curve), which is independent of viewing angle, and always
reduces the observed flux.

\begin{figure*}
\hspace{-0.5in}  \includegraphics[scale=0.32]{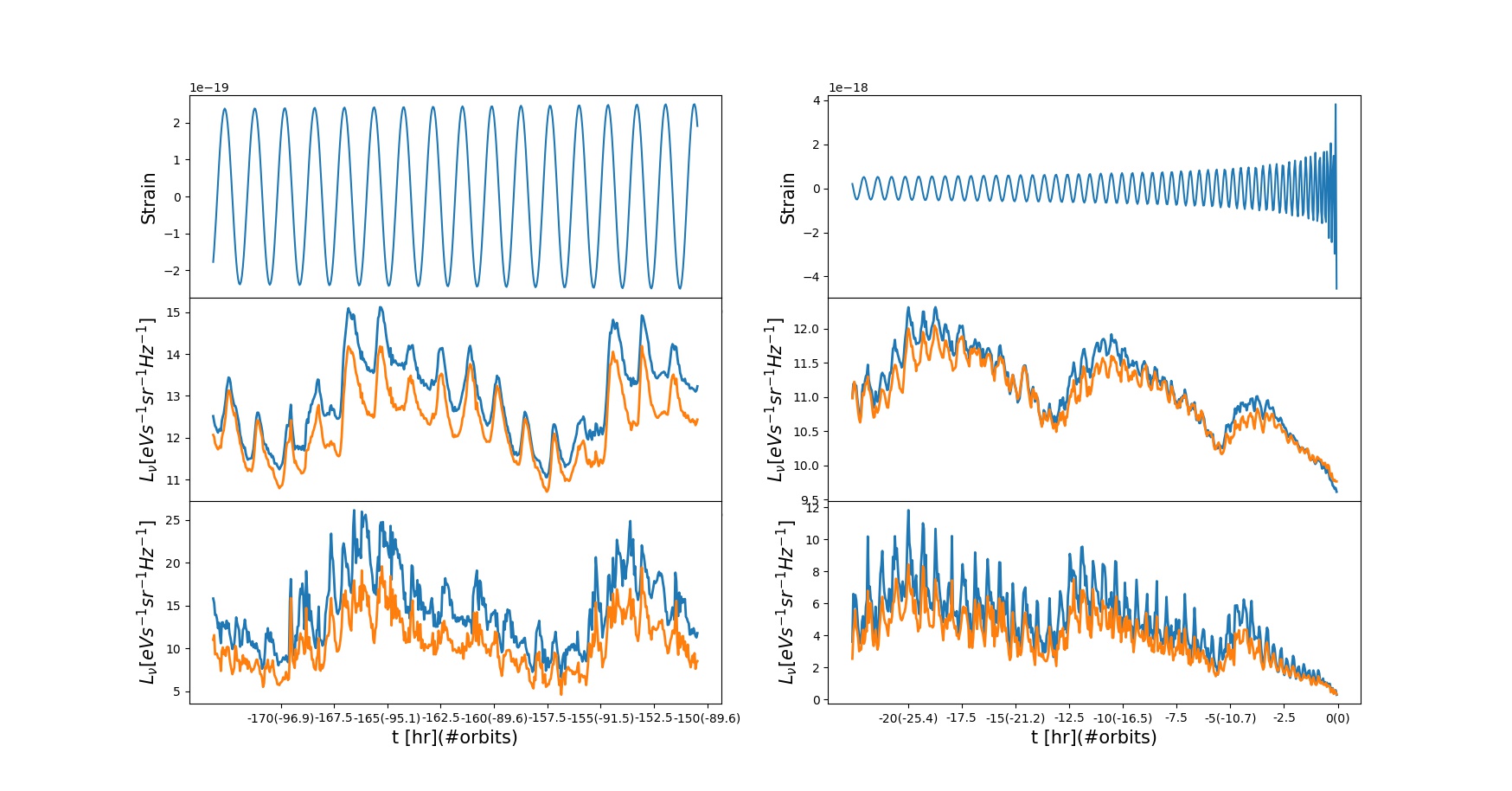}
%
%
\protect\caption{Gravitational wave strain (top panels) and X-ray
  light curves (middle and bottom) calculated at the beginning (left)
  and end (right) of the simulated inspiral in the LISA band. The top
  panel shows the GW strain observed edge-on from a binary at redshift
  $z=1$. The middle panel corresponds to the 2 keV band and bottom
  panel to 10 keV band. The blue curves show the light curves
  neglecting the relativistic Doppler shift or gravitational
  redshift. The orange curves show the light-curves including these
  effects.  While the Doppler effect imposes a small additional
  periodic modulation due to the binary's orbital motion, its dominant
  effect is a nearly uniform dimming of the light-curve (this results
  from the asymmetry between the large blue and redshifts due to the
  internal motion of gas in the minidiscs; see text for discussion).
  The $x$-axis shows the look-back time in the source's rest frame,
  and the number of binary orbits (in parentheses).}
\label{light_curve}
\end{figure*}

In order to study the variability of the spectrum at different
frequencies, we calculate the light curves in the soft (at 2 keV) and
hard (at 10 keV) X-ray bands.  In Figure~\ref{light_curve} we show the
light curve at the beginning of the simulation and right before
merger, with and without the Doppler effect.  In the 2 keV band, we
find clear periodicity in the light curve throughout the inspiral. In
the 10 keV band, the inspiral light curve at the beginning is much
noisier.  We can still see some periodicity by eye, but it is harder
to distinguish than in the 2keV case. However in the late stages of
the inspiral, the 10 keV band also shows clear periodicity at the
orbital period. The light curves also show an overall see-saw
modulation, with a period approximately eight times longer than
$t_{\rm orb}$.  Furthermore, the light curves show a clear a behavior
analogous to the GW chirp signal, in that the periodicity tracks the
binary's increasing orbital frequency.

\begin{figure}
\hspace{-0.1in} \includegraphics[scale=0.4]{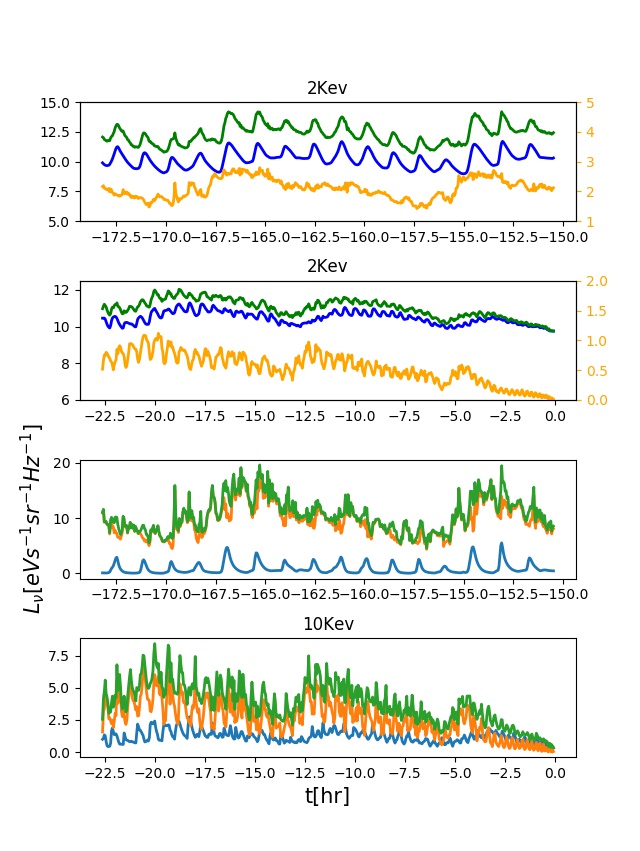}
\protect\caption{Decomposition of the light curves. The orange and
  blue curves correspond to emission arising from $r<a$ (dominated by
  gas in/near the minidiscs) and from $r>a$ (dominated by the
  circumbinary disc).  The green curves correspond to the total
  emission. The upper two panels show the light-curves in the 2 keV
  band, and the bottom two panels in the 10 keV band. For clarity of
  display, the orange curves, which show sub-dominant emission from
  the minidiscs in the soft band in the upper two panels, use the
  $y$-axis labels on the right side.}
\label{light_curve_composition}
\end{figure}

In order to identify the origin of the periodicity in the light
curves, in Figure~\ref{light_curve_composition} we decompose them into
emission arising from $r<a$ (dominated by gas in/near the minidiscs)
and from $r>a$ (dominated by the circumbinary disc).  This figure
reveals that the emission from the circumbinary disc is clearly
periodic from the beginning of the simulation all the way to the
merger in both bands. This is because each BH periodically interacts
with the cavity wall and creates a stream which hits the cavity wall,
shock-heats the gas, and leads to enhanced X-ray emission. By
comparison, the emission from the minidiscs is noisy in the beginning,
and develops clear periodicity only closer to the merger.  One
explanation for this phenomenon is that in the earlier stage the
minidiscs act as buffers of the accretion streams. Unlike steady-state
$\alpha$ discs, the internal shocks in the minidiscs create
significant noise; their overall shape often differs significantly
from a circular disc.  In the late stage of the inspiral (in the last
60 hours before the merger), as the truncation radius decreases, this
buffer effect declines, and the narrow accretion streams eventually
connect the cavity wall and event horizon almost directly (see the
right panel of Fig.~\ref{density_Teff}).  Therefore the noise produced
by the minidiscs gradually disappears.

In the 2keV band, emission from the circumbinary disc dominates, and
the light-curve therefore also shows very clear periodicity, during
the entire inspiral process. The opposite situation occurs in the
10keV band, in which the total emission is dominated by the
minidiscs. The light-curve is noisy in the beginning, and shows much
clearer periodicity in the end stage. From the bottom-most panel of
Fig.~\ref{light_curve_composition}, we see that the emission from the
minidiscs and circumbinary discs have the same period, but with
different phases. Therefore in the later stages, when the strength of
the emission from these two regions are comparable, the total
light-curve sometimes shows two maximas in a single period.

\citet{Haiman2017} and \citet{DanHaiman2017} proposed a toy model for
the minidiscs as two ``light bulbs'', both of which are steady in
their own rest-frame. Therefore, unless observed perfectly face-on,
the Doppler modulation will lead to periodicity at the orbital
frequency.  Note that the light-curve would still be modulated, even
in the case of an equal-mass binary, because the red- and blueshifts
of the thermal emission from the two BHs, with equal but opposite
line-of-sight velocities, will not exactly cancel. However, in our
simulation, we find that the internal structure of the minidiscs are
far from stationary, with shocks forming and streams of gas entering
and leaving the fiducial Hill sphere. The emission from the minidisc
regions is therefore quite noisy. As a result, the Doppler modulation
may be difficult to extract from the observed light-curve. We expect
that this may still be feasible, given a GW template from LISA, with
which the X-ray light-curve can be cross-correlated, but this requires
further analysis. Fortunately, we find that the shock-heated cavity
wall provides a clear periodicity throughout the entire inspiral
process. Moreover, since this periodicity does not originate from the
Doppler effect (see Fig.~\ref{light_curve}), it should exist
independently of the inclination angle.

\begin{figure*}
  \hspace{-0.3in}\includegraphics[scale=0.49]{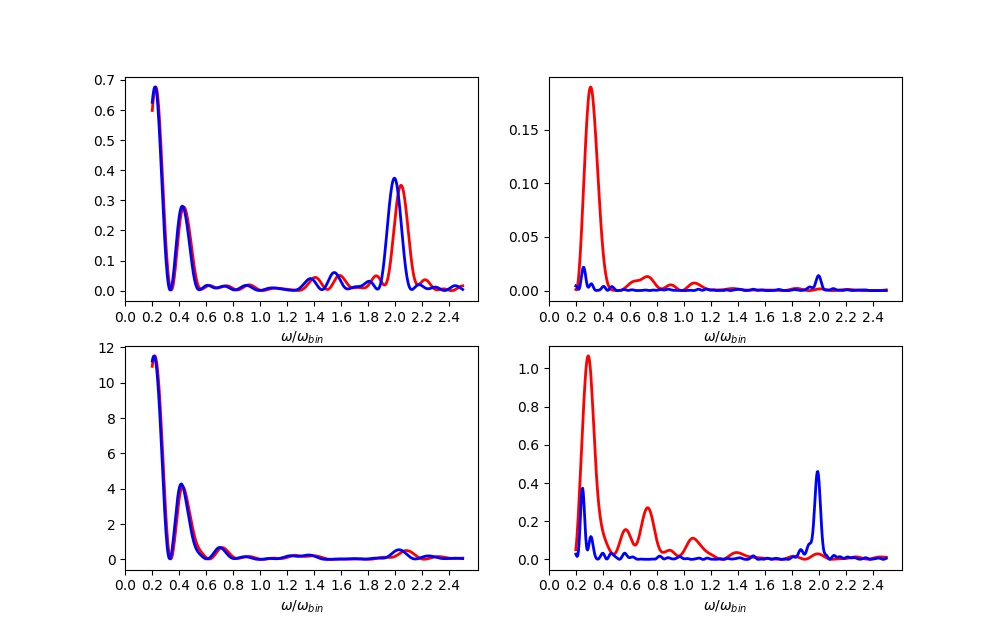}
  \protect\caption{Lomb-Scargle periodograms of the light-curves.  The
    four panels correspond to those in Figure~\ref{light_curve},
    namely to the 2 keV and 10 keV bands (top and bottom,
    respectively) and to the beginning and end of the simulated
    inspiral (left and right, respectively) In each panel, the red
    curve corresponds to the periodogram of the original light-curves
    shown in Figure~\ref{light_curve}. The blue curves correspond to
    scaled light-curves, in which time is measured in units of the
    instantaneous binary orbital period. }
\label{periodogram}
\end{figure*}

In order to assess how well the periodicity in the light-curves track
the binary's orbital period, we rescale the light-curve time-series
with the evolving binary orbital period.  In Figure~\ref{periodogram},
we plot the Lomb-Scargle periodograms with and without this
time-rescaling.  The rescaled periodograms show peaks at $\approx
2\times\Omega_{\rm orb}$, both at the beginning and the end of the
inspiral, implying that this periodicity is able to follow the
accelerating binary orbital frequency (i.e. the ``chirp'') throughout
the entire inspiral process. The first peak in the periodogram without
rescaling arises from a dense lump in the lopsided cavity wall, as
reported in a series of previous works
(e.g. \citealt{2008ApJ...672...83M,Krolik+2010,Noble+2012,Shi+2012,2013MNRAS.436.2997D,2014ApJ...783..134F,2009MNRAS.393.1423C,Roedig2011,Noble+2012,Bowen+2017}).
This feature is smeared out and disappears after time-rescaling,
because it does not follow the binary's frequency chirp.  More
specifically, the frequency of this low-frequency peak increases from
$\approx 0.25\Omega_{\rm orb}(0)$ to $\approx 0.4\Omega_{\rm orb}(0)$,
but this increase is slower than the evolving binary orbital frequency
$\Omega_{\rm orb}(t)$.  Just prior to the merger, the period of this
low-frequency modulation is around 15 orbits. Similarly, in
Fig.~\ref{density_Teff}, we see that the overall size of the cavity is
shrinking, but cannot keep up with the binary separation, which
decreases more rapidly.  Nevertheless, this low-frequency modulation
persists throughout the inspiral process, as we can clearly see in
Figure~\ref{light_curve}. In principle, the evolution of this
low-frequency modulation could be measured in the X-ray light-curves,
and would probe the gas morphology near the central cavity.

\section{Discussion}
\label{sec:discuss}

One of the motivations of this work was to compute the modulation of
the light-curves due to relativistic Doppler effect. Naively, such a
modulation inevitably arises from the same orbital motion that
produces gravitational waves, and therefore the GW and X-ray chirp
signals should track one another with a known phase
\citep{Haiman2017}.  This would be of particular interest for LISA,
allowing a robust measurement of the speed difference of photons and
gravitons, and more generally, a secure identification of the EM
counterpart of a LISA binary.

Our results suggest that it may be difficult to extract this Doppler
modulation from the X-ray light-curve.  We find efficient fueling of
both BHs throught the inspiral process, with gas funneled inside each
BH's Hill radius, forming tidally truncated ``minidiscs''.  These
minidiscs persist nearly all the way to the merger, until the tidal
truncation radius shrink to a size comparable to that of the ISCOs
(roughly 60 hours prior to merger). However, the effective
temperature, density, and velocity fields of the gas inside these
minidiscs are not stationary, and the net Doppler shift is dominated
by these large ``internal'' gas motions, rather than the orbital
velocity of the BHs.  Interestingly, as Figures~\ref{spectra_doppler}
and \ref{light_curve} show, the most prominent effect of the Doppler
shift in the 2-10 keV range is an overall dimming of the ligh-curve,
rather than a sinusoidal modulation.  This is because the hot gas
dominating this emission is located very close to each BH (near or
inside its Hill radius), and the velocity of this gas is larger than
the orbital speed of the BHs.  Regardless of the viewing direction,
there is always gas with both large blue-- and redshifts along the
line of sight, and the total composite spectrum includes both of these
shifts.  The net result can be either an overall dimming or
brightening, depending on the local slope and curvature of the
spectrum.  In other words, for the steeply curved black-body spectral
shape, the Doppler effect is not symmetric under switching the sign of
the line-of-sight velocity.  Figure~\ref{spectra_doppler} shows that
in the 2-10 keV range, the redshift and the corresponding dimming from
receding gas dominates over the blueshift and brightening from
approaching gas.

We emphasize that this result is sensitive to the spectral shape.  One
caveat is that, as mentioned above, due to numerical constrains our
simulated disc is thicker than a typical AGN disc of a solitary BH.
If we were to scale the simulation to correspond to a higher Mach
number, this would shift the normalization of the photon energy.  For
example, the soft vs hard X-ray bands could become FUV and EUV bands
for a disc with a Mach number of $\mathcal{M}_{a} \sim 1000$, instead
of the $\mathcal{M}_{a}=10$ we adopted (recall that $h\nu\propto
\mathcal{M}_{\rm a}^{-5/4}$ for single black hole accretion disc ).
On the other hand, the gas in the inner regions is being heated
primarily by shocks, rather than viscosity, and we find that $h/r$ in
the inner regions of the flow remains $\sim$ 0.2 throughout the
simulation, justifying the scaling to lower Mach numbers (see
Fig.~\ref{hor}).  However it requires further study to understand how
this depends on the choice of $\mathcal{M}_{a}$.

We also note that in eq.~\ref{eq:qcool}, we assumed that the fluid is
gas-pressure dominated; this assumption is likely violated in the
shock-heated streams and minidiscs.

Another caveat in our conclusions is that we focused on thermal
emission.  We have verified that the effective absorptive opacity,
defined as $\tau_{\rm eff} = [\tau_{\rm abs} (\tau_{\rm abs}+\tau_{\rm
    scat})]^{1/2}$, where $\tau_{\rm abs}$ is the optical depth to
free-free absorption and $\tau_{\rm scat}$ is the optical depth to
electron scattering, remains $\tau_{\rm eff} \sim 10^{4} \gg 1$ throughout
our simulation.   This justifies considering a thermalised spectrum.
However, empirically, the X-ray emission from AGN consist of a
combination of thermal disc emission and a hot corona, and in this
paper, we have not included the latter component.  If the emission
from a corona turns out to be more stable (i.e. if the corona lacks
large internal bulk motions), the Doppler modulation in the X-ray band
may be dominated by the BH's orbital motion, and much more prominent.
Moreover, in this paper, we have focused on an equal-mass binary.  We
expect that the Doppler modulation would be more prominent in the case
of low-mass ratio binaries, because the circum--primary and
--secondary minidiscs would be more stable, and there would be less
cancellation between their emission.

The overall shape of the light-curves we found show a clear "X-ray
chirp", caused by the interaction between the binary and the gas
discs.  The periodicity of the chirp tracks the shrinking binary
orbital period $t_{\rm orb}$, in tandem with the GW signal.  Since the
chirp is produced by hydrodynamical modulation, its absolute phase is
not tied directly to the GW emission.  Nevertheless, the speed
difference between photons and gravitons should still be measurable,
in principle, given a precise measurement of the phase-evolution of
both the GW and the X-ray chirp signals.  In the future, it would be
interesting to compute the accuracy to which the phase difference can
be measured in practice, if the absolute phase is unknown ab-initio,
given realistic observational S/N from LISA, and from an X-ray
instrument such as Athena\footnote{see
  \texttt{www.cosmos.esa.int/web/athena}}, Lynx\footnote{See
  \texttt{www.astro.msfc.nasa.gov/lynx}}, ULTRASAT\footnote{See
  \texttt{www.weizmann.ac.il/ultrasat}} or EXTP~\citep{extp}.
In this paper we neglect spin of the BHs, but the spin would change
the GW signal significantly, and would further complicate a direct
comparison of the X-ray lightcurve with the LISA signal.

We have made several simplifying assumptions throughout this paper,
which we intend to relax in our future work.  We have focused here on
an equal-mass binary, but we intend to study the dependence of the
spectra and light-curves on the binary mass ratio.  Similarly, we
intend to determine the dependence of these features on the disc
thickness. We also used an $\alpha$--prescription for the viscosity.
We intend to perform magnetohydrodynamic (MHD) simulations in the
future (e.g. \citealt{Krolik+2010}).  In this work, we have also
assumed that the binary BHs are on circular orbits, with zero
eccentricity.  On the other hand, eccentricity is likely to develop as
a result of strong gas torques during the earlier stages of the
inspiral
(e.g.~\citealt{2009MNRAS.393.1423C,Roedig2011,2012A&A...545A.127R,2017MNRAS.466.1170M,Sesana2011}).
In the future, we intend to study live binaries, and follow the
development of eccentricity, and to asssess the impact of eccentricity
on the accretion rates and the predicted spectra and
light-curves.

In the present simulation, we followed a pseudo-Newtonian approach; a
full relativistic simulation of the gas in the vicinity of the event
horizon could modify our results.  Finally, we started our current
simulations past the fiducial 'decoupling' stage. In the cartoon
picture of the binary inspiral \citep{Milosavljevic2005}, once the
binary is sufficiently compact that the GW inspiral timescale is
shorter than the viscous timescale in the nearby disc, the BHs outpace
the disc and leave the disc gas behind.  In this picture, our
procedure of holding the binary's orbit initially fixed, and allowing
it to reach a steady state, would not be justified.  However, from the
results we find, as well as from our earlier work \citep{Farris15}, we
see that the cartoon picture is inaccurate, and the cavity can, in
fact, shrink and follow the inspiraling binary, even well past the
nominal decoupling.  We therefore expect our way of setting up the
initial condition to be reasonably accurate, but we intend to study
this in future work, by beginning simulation near or prior to the
decoupling stage.

\section{Conclusions}
\label{sec:conclude}

We have performed 2D simulations of accretion onto an inspiraling and
merging binary BH system, and studied the corresponding EM signatures.
We have computed the evolving multi-color black body emission from
snapshots of our simulations, including the effects of the
relativistic Doppler shift and gravitational redshift.  The most
important conclusions of this work can be summarized as follows:

(1) {\em The light-curves show clear periodicity}.  Accretion onto the
individual BHs has long been known to be periodic~\citep{al94}, and it
is not surprising that any luminosity produced by gas near the
individual BHs is periodic.  However, we have found that the emission
arising from gas farther out, near the cavity wall, which dominates
the softer X-ray bands, shows a similar (and even clearer)
periodicity. This is because streams of gas, flung outwards by the
BHs, periodically hit and shock the cavity wall.  The frequency is
twice the binary's instantaneous orbital frequency.

(2) {\em Distinct behaviour in different bands.} In the beginning of
the simulation, the emission from the minidiscs is noisy, but develops
clear periodicity during the later stages. In the soft X-ray band, the
emission from the circumbinary disc dominates, and the light curve
shows a clear periodicity from the beginning of the simulation
($a=60GM/c^{2}$). By comparison, in the hard X-ray band, the emission
in the beginning is dominated by the minidiscs, and periodicity in the
early-inspiral light curve is obscured by noise, with clearer
periodicity developing only in the late stages.

(3) {\em Doppler modulation is sub-dominant.} We find that the Doppler
modulation of the light-curve does not follow a simple toy model of
two moving light-bulbs in orbit.  This is because the effective
temperature, density, and velocity field of the gas in the minidiscs
are not stationary, and the line-of-sight velocity of this gas is
dominated by the large internal gas motions inside the minidiscs,
rather than the clean orbital motion of the BHs.  We find that the
most conspicuous effect of the Doppler shift is an overall dimming
(1keV$\lsim h\nu\lsim 20$keV) or brightening ($h\nu\gsim 20$keV) of
the light-curves.  We caution that these results are sensitive to the
spectral shape and emission mechanism. In particular the Doppler
effect could more closely track the binary's orbital motion for X-ray
emission from more stable coronae around the individual BHs which do
not have large internal motions.

(4) {\em The EM chirp follows the GW inspiral.} We have found that the
minidiscs persist until the very end of the inspiral process.  The
discs are truncated at the gradually shrinking size of the Hill
radius.  The thermal emission from the minidiscs declines gradually,
but remains overall significant all the way to the merger.  The
overall periodicities in both the soft and hard X-ray bands track the
shrinking orbital period of the binary. 

(5) {\em No ``second decoupling''}.  \citet{2017MNRAS.468L..50F}
suggested that the inner (circum-primary) disc will become
geometrically thick ($h/r \sim$ 1) in the late stages of the inspiral,
as a result of tidal heating. They find, using a 1D analytical models
and numerical simulation, that the viscous time in the inner disc
decreases to $t_{\rm vis}\sim t_{\rm GW}$, and it effectively
decouples from the binary.  By comparison, in our 2D simulation, the
size and surface density of the two minidiscs are continuously
adjusted, and eventually, in the late stages, the minidiscs disappear,
and narrow accretion streams instead fall inside the event horizon
directly.  The inspiral process does thicken the minidiscs, but the
aspect ratios remain below $h/r \sim 0.2$.

\vspace{\baselineskip}

The simulations presented here are based on many simplifications.
Nevertheless, our results suggest that an EM chirp signal may
accompany the GW emission, with the EM period tracking that of the GW
signal right up to the merger of a SMBH binary.  A detection of this
EM chirp could provide a potential ``smoking gun'' evidence to
uniquely identify the EM counterpart of a SMBH binary detected by
LISA.

\section*{Acknowledgements}

We thank Daniel D'Orazio, Paul Duffell, Julian Krolik and Geoffrey
Ryan for useful discussions.  Financial support was provided by NASA
through ATP grant NNX15AB19G and ADAP grant NNX17AL82G and by NSF
grants 1715661 and 1715356.  ZH also gratefully acknowledges support
from a Simons Fellowship in Theoretical Physics and hospitality by NYU
during his sabbatical leave when this work began. This work was
performed in part at Aspen Center for Physics, which is supported by
National Science Foundation grant PHY-1607611.





\bibliography{paper}
\bibliographystyle{mnras}

\label{lastpage}

\end{document}